\documentclass[onecolumn,secnumarabic,amssymb, amsmath, nofootinbib,tightenlines,
nobibnotes, aps, prl,epsfig]{revtex4}
\usepackage{graphicx}
\usepackage{dcolumn}
\usepackage{bm}
\begin{document}
\preprint{APS/123-QED}
\title{A simple model for the charm structure
function of nuclei
 }

\author{G.R.Boroun}%
 \email{boroun@razi.ac.ir }
\affiliation{ Department of Physics, Razi University, Kermanshah
67149, Iran}
\date{\today}
\begin{abstract}
In this paper, we have investigated the importance of quark charm
in nuclear structure functions in the color dipole model at small
$x$. The charm structure function per nucleon $F_{2}^{cA}/A$ for
light and heavy nuclei in a wide range of transverse separations
$\mathrm{r}$ with renormalization and factorization scales are
considered. Bounds on the ratio $F^{cA}_{2}/AF^{A}_{2}$ for nuclei
are well described with respect to the electron-ion future
colliders kinematic range, i.e, EIC
and EIcC colliders.\\

\end{abstract}
 \pacs{***}
\maketitle
\subsection{I. Introduction}

Electron-nucleus scattering experiments provide vital
complementary information to test, assess and validate different
nuclear models and data collected from the electron scattering
controlled kinematics, large statistics and high precision allow
one to constrain nuclear properties and specific interaction
processes [1]. The study of nuclear structure function and its
modification at small Bjorken $x$, is a very interesting subject
in compared to those for free nucleon observed in deep inelastic
scattering (DIS) [2]. For a nucleus $A$ with $Z$ protons and
$N=A-Z$ neutrons, the nuclear parton distribution functions
(nPDFs) are defined into the parton distribution functions (PDFs)
of a bound proton and neutron (i.e., $f_{i}^{p/A}$ and
$f_{i}^{n/A}$ respectively) by the following form
\begin{eqnarray}
f_{i}^{A}(x,Q^2)=\frac{Z}{A}f_{i}^{p/A}(x,Q^2)+\frac{N}{A}f_{i}^{n/A}(x,Q^2),
\end{eqnarray}
where, the bound nucleon PDFs are different from those of a free
proton by the nuclear modification factors as
\begin{eqnarray}
R_{i}^{A}(x,Q^2)=\frac{f_{i}^{p/A}(x,Q^2)}{f_{i}^{p}(x,Q^2)}.
\end{eqnarray}
The modification of nuclear structure functions at small Bjorken
$x$ in comparison with the free nucleon observed in DIS is defined
by small-$x$ shadowing followed by antishadowing and EMC-effect,
and at the large-$x$ is Fermi motion. At small values of the
Bjorken variable $x$ (for $x<0.01$), the shadowing effect is seen
in nuclear DIS. In the shadowing region, the structure function
$F_{2}$ per nucleon turns out to be smaller in nuclei than in a
free nucleon [3]. This effect manifests itself as an inequality
$F_{2}^{A}/(AF_{2}^{N})<1$, where $A$ is the number of nucleons in
a nuclear target. Indeed, unitarity driven nuclear shadowing
becomes important at $x{\ll}x_{A}=1/(m_{N}R_{A})=0.15A^{-1/3}$
where $R_{A}$ is the radius of the target nucleus and $m_{N}$ is
the nucleon mass [4]\footnote{For further discussion refer to the
Refs.[5-8]}. Nuclear shadowing, in the laboratory frame, derives
from the coherent interaction of $q\overline{q}$,
$q\overline{q}g$,... states. The Fock state expansion of the
physical virtual photon reads
$|\gamma^{*}>=\sqrt{z_{g}}\Psi_{q\overline{q}}|q\overline{q}>+\Phi_{q\overline{q}g}
|q\overline{q}g>$, where $\Psi_{q\overline{q}}$ and
$\Phi_{q\overline{q}g}$ are the light-cone wavefunctions of the
$q\overline{q}$ and $q\overline{q}g$ states. Here $\sqrt{z_{g}}$
is the renormalization of the $q\overline{q}g$ state by the
virtual radiative corrections for the $q\overline{q}g$ state. For
the lowest $|q\overline{q}>$ Fock component of the photon, the
interaction of $q\overline{q}$ dipole of transverse separation
$\mathbf{r}$ with a nucleon represents with the dipole cross
section $\sigma_{q{\overline{q}}}(r)$ [5,6]. Indeed, the incoming
virtual photon splits into a colorless $q\overline{q}$ pair long
before reaching the nucleus, and this dipole interacts with
typical hadronic cross sections which results in absorption.\\
The key feature is the connection of the dipole-target amplitude
to the integrated gluon density where, at very low $x$, the parton
saturation models illuminate the behavior of the gluon density.
The dipole cross section can be derived from the
Balitsky-Kovchegov (BK) equation [9,10], which established a
non-linear evolution equation to describe the high energy
scattering of a $q\overline{q}$ dipole on a target in the fixed
coupling case based on the concept of saturation. For the study of
saturation of nuclei, those have an advantage over protons since
they have more gluons to start with. Therefore, non-linear effects
in the evolution of the nuclear gluon distribution should set in
at much lower energy than for protons. DIS on heavy nuclei is
expected to probe the color dipole cross section in a way
different from DIS on nucleons. Specifically, the larger the
dipole size is the stronger nuclear screening [11-15]. Unitarity
constraints for deep-inelastic scattering on nuclei predicted in
Ref.[16]. Evidently, the nuclear shadowing (screening) depends on
high mass diffraction.\\
It is expected that measurements over the extended $x$ and $Q^{2}$
ranges, which would become possible for future experiments, in
particular, for experiments at the Electron-Ion Collider (EIC)
[17] and Electron-Ion Collider in China (EIcC) [18], will give
more information in order to discriminate between the distinct
models of shadowing of the QCD dynamics at small $x$. These future
facilities will probe nuclear structure over a broad range of $x$
and $Q^2$ and the analysis of the nuclear effects in deep
inelastic scattering has been a topic of discussion in the
community in recent times \footnote{The center-of-mass energies in
electron-ion colliders proposed in China and the US are
$15-20~\mathrm{GeV}$ for EIcC and $30-140~\mathrm{GeV}$ for EIC
respectively. }. One of the main physics goals of these future QCD
laboratories at small $x$ will be to unambiguously unveil the
onset of the so-called gluon saturation regime of QCD, which is
characterized by a transverse momentum scale, the saturation scale
$Q_{s}(x)$, at which non-linearities become of comparable
importance to linear evolution. In the last years the analysis of
the nuclear effects in deep inelastic scattering (DIS) has been
extensively discussed in the literature [19-31]. In this paper
there is a good chance to produce interesting new predictions for
the charm structure function of nuclei for future experiments in
the low $x$ region. These calculations are based on the color
dipole picture (CDP) with a characteristic saturation momentum,
$Q_{s}^{A}$. We analyze the charm quark structure functions in
nuclei and those ratios in a wide range of $r$ in section II. In
this section, the heavy quark structure functions can be combined
with the Sudakov form
factor. Section III contains our results and conclusions.\\

\subsection{II. Method}

The cross section in the dipole formulation of the photon-nucleon
scattering is defined, with respect to the polarization
(transverse, T, or longitudinal, L) of the virtual photon, by
\begin{eqnarray}
\sigma_{L,T}^{\gamma^{*}p}(x,Q^{2})=\int dz d^{2}\mathbf{r}
|\Psi_{L,T}(\mathbf{r},z,Q^{2})|^{2}\sigma_{\mathrm{dip}}({x},\mathbf{r}),
\end{eqnarray}
where $\Psi_{L,T}$ is the corresponding photon wave function in
mixed representation and $\sigma_{\mathrm{dip}}({x},r)$ is the
dipole cross-section which related to the imaginary part of the
$(q\overline{q})p$ forward scattering amplitude. The transverse
dipole size $r$ and the longitudinal momentum fraction $z$ due to
the photon momentum are defined.  The variable $z$, with $0\leq z
\leq 1 $, characterizes the distribution of the momenta between
quark and antiquark. The Golec-Biernat and Wusthoff (GBW) model
[32] is a model for the so-called dipole cross section, that is
used to descrobed the inclusive DIS data for $x<0.01$ and all
$Q^2$. The model reads
\begin{eqnarray}
\sigma_{\mathrm{dip}}(x,r)=\sigma_{0}\bigg{\{}1-\exp\bigg{[}
\frac{1}{4}Q^{2}_{\mathrm{sat}}r^2\bigg{]}\bigg{\}},
\end{eqnarray}
where $Q_{\mathrm{sat}}(x)$ is the saturation scale defined as
$Q^{2}_{\mathrm{sat}}(x)=Q_{0}^{2}(x_{0}/x)^{\lambda}$ for the
proton. Three parameters, $\sigma_{0}$, $\lambda$ and $x_{0}$,
were determined [33] from a fit to the HERA data and have values,
$\sigma_{0}=27.32 \pm 0.35~ \mathrm{mb}$, $\lambda=0.248 \pm
0.002$ and $x_{0}/10^{-4}=0.42 \pm 0.04$ for the 4-flavor
respectively. The fixed parameters are
$Q_{0}^{2}=1~\mathrm{GeV}^2$,
$m_{u}=m_{d}=m_{s}=0.14~\mathrm{GeV}$ and
$m_{c}=1.4~\mathrm{GeV}$. The quantity $x$ used in expressions
above is the modified Bjorken variable,
$x=x_{\mathrm{Bj}}(1+\frac{4m^2_{q}}{Q^2})$ , with $m_{q}$ being
the effective quark mass. This replacing is a simple way to
regular the divergence of the cross section. This modification is
quite important as heavy quarks contribution are taken into
account and implies that the value of the quark mass plays an
important role in avoiding the divergence of the cross section.\\
The Bartels, Golec-Biernat and Kowalski (BGK) model [34], is
another phenomenological approach to the color dipole cross
section and reads
\begin{eqnarray}
\sigma_{\mathrm{dip}}(x,\mathbf{r})=\sigma_{0}\bigg{\{}1-
\exp\bigg{(}
\frac{\pi^2r^2\alpha_{s}(\mu^2)xg(x,\mu^2)}{3\sigma_{0}} \bigg{)}
\bigg{\}}.
\end{eqnarray}
The scale $\mu^2$ is connected to the size of the dipole and takes
the form $\mu^2=\frac{C}{r^2}+\mu^2_{0}$, where the parameters $C$
and $\mu_{0}$ are determined from a fit to DIS data [33]. Here
$g(x,\mu^2)$ is the gluon collinear PDF. In the color transparency
domain, $r{\rightarrow}0$, the dipole cross section is related to
the gluon density by the following form [35]
\begin{eqnarray}
\sigma_{\mathrm{dip}}(x,\mathbf{r}){\simeq}
\frac{\pi^2r^2\alpha_{s}(\mu^2)xg(x,\mu^2)}{3}.
\end{eqnarray}
The gluon distribution in the GBW and BGK  models take the form
\begin{eqnarray}
xg(x,\mu^2)=\frac{3\sigma_{0}}{4\pi^2\alpha_{s}(\mu^2)}Q_{s}^2
\end{eqnarray}
where $\alpha_{s}$ is the running coupling at $\mu^2$ scale [36].
The expression for the nuclear gluon distribution
$xg^{A}(x,\mu^{2})$ is the same expect for the change
$Q_{s}^2{\rightarrow}Q_{s}^{2A}$ with the replacement the area of
the target with the coefficient $A^{2/3}$. Therefore, the gluon
distribution for a nuclear target with the mass number A is
defined by
\begin{eqnarray}
xg^{A}(x,\mu^2)=\frac{3\sigma_{0}A^{2/3}}{4\pi^2\alpha_{s}(\mu^2)}Q_{s}^{2A},
\end{eqnarray}
where
\begin{eqnarray}
Q_{s}^{2A}=Q_{s}^{2}\bigg{(}  \frac{A\pi R_{p}^{2}}{\pi R_{A}^{2}}
\bigg{)}^{\frac{1}{\delta}}
\end{eqnarray}
In Ref.[25], it was found $\delta=0.79{\pm}0.02$ and the nuclear
radius is given by the usual parameterization
$R_{A}=(1.12A^{1/3}-0.86A^{-1/3})~\mathrm{fm}$ and $\pi
R^{2}_{p}=1.55{\pm}0.02~\mathrm{fm}^2$. The charm structure
function in nuclei, owing to the dominance of the gluon
distribution, in the collinear generalized double asymptotic
scaling (DAS) approach [37] is defined by the following form in
the small $x$ region as
\begin{eqnarray}
F_{2}^{cA}(x,\mu_{r}^{2}){\simeq}~e^{2}_{c}\sum_{n=0}(\frac{\alpha_{s}}{4\pi})^{n+1}B^{(n)}_{2,g}(x,\xi_{r})
{\otimes} xg^{A}(x,\mu_{r}^{2}),
\end{eqnarray}
where $B_{2,g}$ is the collinear Wilson coefficient function in
the high energy regime [38] and $n$ denotes the order in running
coupling $\alpha_{s}$. Here, $e^{2}_{c}$ is the squared charge of
the charm and $\xi_{r}=\frac{m_{c}^{2}}{\mu_{r}^{2}}$.
 The default
renormalisation and factorization scales are set to be equal
$\mu_{R}^{2}=\mu_{r}^{2}+4m_{c}^2$ and
$\mu_{F}^{2}=\mu_{r}^{2}$.\\
In addition, we consider bounds for $F_{2}^{cA}/F^{A}_{2}$ follows
from the standard nuclear dipole picture which gives correlated
values in estimate $F^{A}_{L}/F^{A}_{2}$ into the higher Fock
components of the photon wave function. The nuclear structure
function $F^{A}_{2}$ can be obtained from the $\gamma^{*}A$ cross
section through the relation
\begin{eqnarray}
F_{2}^{A}=Q^2\sigma^{\gamma^{*}A}/(4\pi^2\alpha),
\end{eqnarray}
where the nuclear cross section is related to the proton cross
section by the following form
\begin{eqnarray}
\sigma^{\gamma^{*}A}\bigg{(}\frac{Q^2}{Q^{2A}_{s}}\bigg{)}=
\bigg{(}\frac{\pi R_{A}^{2}}{\pi R^{2}_{p}}\bigg{)}
\sigma^{\gamma^{*}p}\bigg{(}\frac{Q^2}{Q^{2A}_{s}}\bigg{)}
\end{eqnarray}
where the $\gamma^{*}p$ cross section reads [25,28]
\begin{eqnarray}
\sigma^{\gamma^{*}p}=
\overline{\sigma}_{0}\bigg{[}\gamma_{E}+\Gamma\bigg{(}0,\frac{a(Q^{2A}_{s})^{b}}{(Q^2)^{b}}\bigg{)}
+{\ln}\bigg{(}\frac{a(Q^{2A}_{s})^{b}}{(Q^2)^{b}}\bigg{)}\bigg{]}.
\end{eqnarray}
Here $\gamma_{E}$ and $\Gamma(0,\eta)$ are the Euler constant and
the incomplete $\Gamma$ function respectively, where the fit
parameters (i.e., $a$ and $b$) are $a=1.868$ and $b=0.746$.\\
The bound  value of $F^{cA}_{2}/F^{A}_{2}$ is obtained by the
following form
\begin{eqnarray}
\frac{F_{2}^{cA}}{AF_{2}^{A}}=\frac{4\pi^2{\alpha}e^{2}_{c}}{Q^2}A^{(\frac{1}{\delta}-\frac{1}{3})}
\bigg{(}\frac{\pi R_{p}^2 }{\pi R_{A}^{2}}
\bigg{)}^{(\frac{1}{\delta}-1)}
\frac{\sum_{n=0}(\frac{\alpha_{s}}{4\pi})^{n+1}B^{(n)}_{2,g}(x,\xi_{r})
{\otimes}
xg(x,\mu_{r}^{2})}{\overline{\sigma}_{0}\bigg{[}\gamma_{E}+\Gamma\bigg{(}0,\frac{a(Q^{2A}_{s})^{b}}{(Q^2)^{b}}\bigg{)}
+{\ln}\bigg{(}\frac{a(Q^{2A}_{s})^{b}}{(Q^2)^{b}}\bigg{)}\bigg{]}}
\end{eqnarray}
which will be interesting in EIC and EIcC colliders in the future
energy range. Especially at the EIC, it is expected to be probed
at an essentially low $x$ (up to $x{\sim}10^{-4}$), thus providing
us with new information on the charm quark density in a nuclei.
Indeed, both EIC in the small-$x$ region and EIcC at moderate $x$
give us nuclear modification of the structure functions and hadron
production in deep inelastic scattering eA collisions and new
information on the parton distribution in nuclei [38].\\


\subsection{III. Numerical Results}

In the present paper we consider the charm structure function of
the deep inelastic scattering of nuclei, which is directly related
with the gluon distribution of nuclei in the CDP approach at low
$x$. In this model, the ratios $xg^{A}(x,\mu^2)/Axg(x,\mu^2)$ and
$F_{2}^{cA}(x,\mu_{r}^{2})/AF_{2}^{c}(x,\mu_{r}^{2})$ are
independent of the variables and depend on  the mass number A by
the following form
\begin{eqnarray}
R^{A}{\equiv}\frac{F_{2}^{cA}(x,\mu_{r}^{2})}{AF_{2}^{c}(x,\mu_{r}^{2})}{\propto}\frac{xg^{A}(x,\mu^2)}{Axg(x,\mu^2)}=A^{-1/3}\bigg{(}
\frac{A\pi R_{p}^{2}}{\pi R_{A}^{2}} \bigg{)}^{\frac{1}{\delta}}
\end{eqnarray}
which gives a plateau behavior in the region $x{\leq}0.01$, which
is similar with the results $F_{2}^{A}/AF_{2}$ in Refs.[25,28]. In
Fig.1, we plot this ratio for a wide range of A and observe that
this ratio is independent of $x$ and $r$. We observe that the
ratio of $R^{A}$ rapid drop for $A<50$ followed by a slow rise for
larger $A$. The minimum values of the ratio are found around
$A{\approx}56$ where these nuclei are the most tightly bound [39].
The increase of binding energy to $A{\approx}56$ decreases the
momentum carried by parton distributions in comparison with other
nuclei.\\
\begin{figure}[h]
\includegraphics[width=0.55\textwidth]{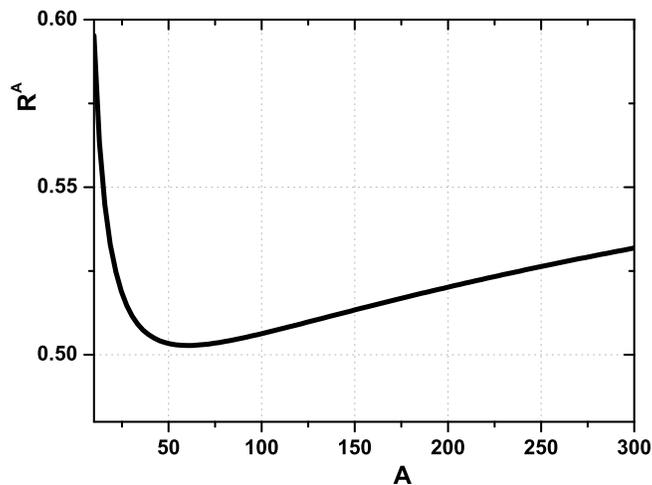}
\caption{The ratios $xg^{A}(x,\mu^2)/Axg(x,\mu^2)$ and
$F_{2}^{cA}(x,\mu_{r}^{2})/AF_{2}^{c}(x,\mu_{r}^{2})$ into the
mass number A. }\label{Fig1}
\end{figure}
Our numerical results for charm structure functions of nuclei per
nucleon, $F_{2}^{cA}/A$  are shown in Fig.2 into  the
uncertainties of the renormalization and factorization scales
\begin{figure}[h]
\includegraphics[width=0.7\textwidth]{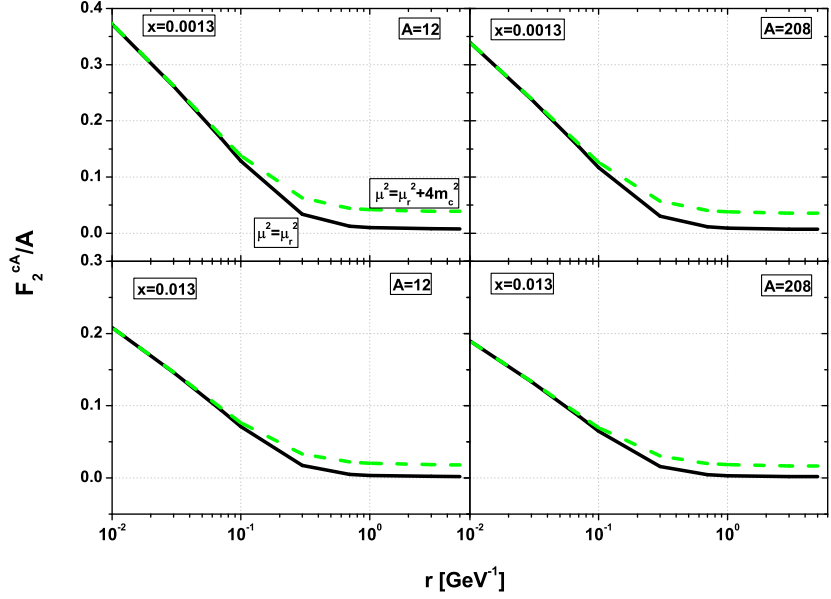}
\caption{Results of the charm structure function per nucleon
$F_{2}^{cA}/A$ for light and heavy nuclei in a wide range of the
transverse separation $\mathrm{r}[\mathrm{GeV}^{-1}]$ with
$x=0.0013$ and $x=0.0130$. The uncertainties are due to
$\mu^{2}=\mu_{r}^{2}+4m_{c}^{2}$ (dashed lines) and
$\mu^{2}=\mu_{r}^{2}$ (solid lines).}\label{Fig3}
\end{figure}
\begin{table}[h]
\centering \caption{The transverse separation range of $r$ in the
 future facilities(i.e., EIcC and EIC) with the
inelasticity $y{\leq}1$ for $x=0.0013$ and 0.0130.
  }\label{table:table1}
\begin{minipage}{\linewidth}
\renewcommand{\thefootnote}{\thempfootnote}
\centering
\begin{tabular}{|l|c||c|c|c||} \hline\noalign{\smallskip} Collider &
$\sqrt{s_{\mathrm{max}.}}$[GeV] & x=0.0013 & x=0.0130 \\
\hline\noalign{\smallskip}
EIC & 140 & r$>$0.1 & r$>$0.03  \\
\hline\noalign{\smallskip}
EIcC & 20 & ---- & r$>$0.3  \\
\hline\noalign{\smallskip}
\end{tabular}
\end{minipage}
\end{table}
in a wide range of $r$ for $x=0.0013$ and $0.0130$. These results
for $F_{2}^{cA}/A$, in Fig.2, increase as $r$ decreases for light
and heavy nuclei. We observe that the results of $F_{2}^{cA}/A $
for light and heavy nuclei, at very low $r$, will increase at the
EIC according to the kinematic coverage of the deep inelastic
scattering process. The uncertainties due to the renormalization
and factorization scales increase as $r$ increase. Owing to the
$x-Q^2$ EIC kinematics \footnote{Please see Fig.1.7 in Ref.[40].},
the new information on the charm structure function in nuclei can
be achieved with $x=0.0130$ for $r{\gtrsim}0.06~\mathrm{GeV}^{-1}$
and with $x=0.0013$ for $r{\gtrsim}0.2~\mathrm{GeV}^{-1}$. As a
result, we predict that at low $r$, the charm structure function
will increases at the EIC than the EIcC at high
inelasticity according to Table I.\\
 \begin{figure}[h]
\includegraphics[width=0.55\textwidth]{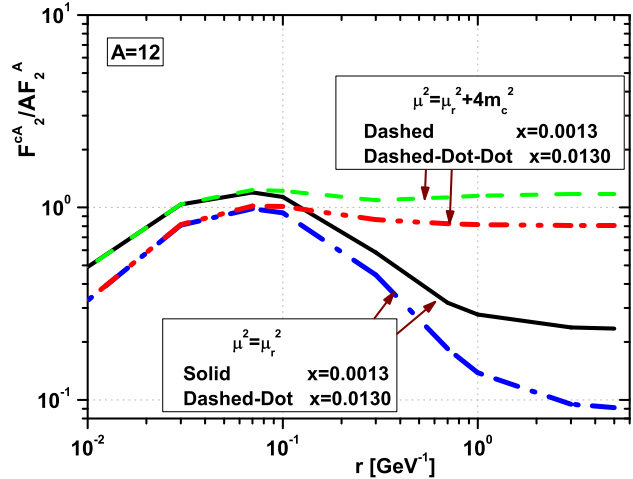}
\caption{ $F_{2}^{cA}/AF_{2}^{A}$ evaluated as a function of $r$
with $\mu^{2}=\mu_{r}^{2}+4m_{c}^{2}$  and $\mu^{2}=\mu_{r}^{2}$
for nuclei A=12 at $x=0.0013$ and $x=0.0130$.}\label{Fig3}
\end{figure}
\begin{figure}[h]
\includegraphics[width=0.55\textwidth]{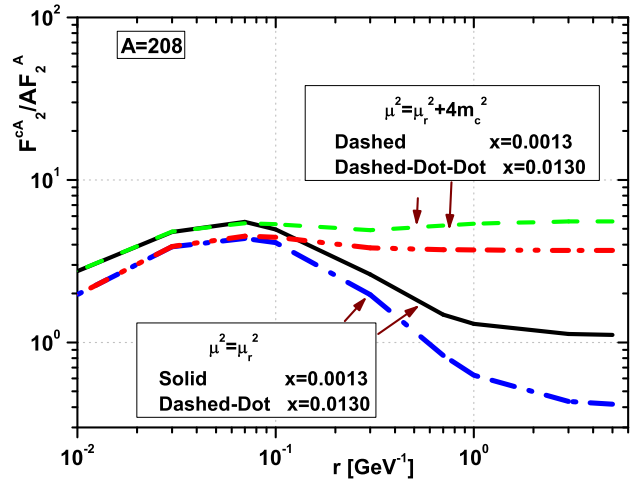}
\caption{The same as Fig.3 for A=208.}\label{Fig4}
\end{figure}
In Figs.3 and 4, we plot the ratio $F_{2}^{cA}/AF^{A}_{2}$ for
nuclei $A=12$ and $A=208$, respectively, as a function of $r$ with
the renormalization and factorization scales with $x=0.0013$ and
$x=0.0130$. These results in Figs.3 and 4 are shown a flat
behavior of $F_{2}^{cA}/AF_{2}^{A}$ with
$\mu^2=\mu_{r}^2+4m_{c}^2$ and decrease sharply with
$\mu^2=\mu_{r}^2$ at $r{\gtrsim}10^{-1}~\mathrm{GeV}^{-1}$. The
renormalization and factorization scales results for light and
heavy nuclei are compatible at $r<10^{-1}~\mathrm{GeV}^{-1}$  and
have the largest uncertainties at $r>10^{-1}~\mathrm{GeV}^{-1}$.
The maximum value of $F_{2}^{cA}/AF_{2}^{A}$, in accordance to the
EIC kinematic range, is $\simeq 1.3$ and $5.5$ for nuclei $A=12$
and $A=208$ respectively. Indeed, the importance of the nuclear
structure function ratios will depend on the values of
$F_{2}^{cA}/AF_{2}^{A}$, where these bounds can further restrict
the kinematical range of the applicability of the dipone picture
in the future electron-ion colliders (i.e., EIC and EIcC).\\

Summarizing, a simple model for the charm structure functions in
nuclei, in the region of small $x$,  has been presented. We
analyzed $F^{cA}_{2}$ using the gluon density from the GBW and BGK
models, inspired by the DAS approach, within the color dipole
model to the future electron-ion  colliders kinematic range at EIC
and EIcC, in a wide range of transverse separations $\mathrm{r}$.
Our results indicate that the study of the charm structure
functions in the eA process at EIC is ideal for considering the
heavy quark effects present in the nuclear structure functions,
which, in turn, is a crucial ingredient to estimate the bounds of
the processes which will be studied in future accelerators. We
have considered the charm structure function $F_{2}^{cA}/A$ per
nucleon in light and heavy nuclei, then have obtained bounds on
$F_{2}^{cA}/AF_{2}^{A}$ at moderate and large $r$ with the
renormalization and factorization scales. We demonstrated the
importance of the contributions of $F_{2}^{cA}/A$ and
$F_{2}^{cA}/AF_{2}^{A}$ at small $r$ in the EIC and EIcC
colliders. The uncertainties of these results are due to  the
standard variations in the renormalization and factorization
scales which increase as $r$ increases. The focus of this paper
was to provide an analytical charm structure function per nucleon
in nuclei for studying high energy lepton-nucleus phenomena at
future colliders such as EIC and the EIcC.\\

\subsection{ACKNOWLEDGMENTS}
The author is grateful to Razi University for the financial
 support of this project. Thanks is due to N.N.Nikolaev for useful discussions.\\



\section{References}

1. V.Pandey, Phys.Sci.Forum {\bf 8}, 1 (2023).\\
2. European Muon, J. Aubert et al., Phys.Lett.B {\bf123}, 275
(1983).\\
3. P.Paakkinen, arXiv [hep-ph]: 1802.05927.\\
4. N.N.Nikolaev and V.I.Zakharov, Phys.Lett.B {\bf55B}, 397
(1975); N.N.Nikolaev, W.Schafer, B.G.Zakharov and V.R.Zoller,
J.Exp.Theor.Phys.{\bf97}, 441 (2003).\\
5. M. Krelina and J.Nemchik, Eur.Phys.J.Plus {\bf135}, 444
(2020).\\
6. N.N.Nikolaev, W.Schafer, B.G.Zakharov and V.R.Zoller,
J.Exp.Theor.Phys.Letters {\bf84}, 537 (2006).\\
7. J.L.Albacete, N.Armesto, A.Capella, A.B.Kaidalov and
C.A.Salgado, arXiv[hep-ph]:0308050, Report number:
CERN-TH/2003-184.\\
8. L.S.Moriggi, G.M.Peccini and M.V.T.Machado, Phys.Rev.D
{\bf103}, 034025 (2021).\\
9. Y.V.Kovchegov, Phys.Rev.D {\bf60}, 034008 (1999);  Phys.Rev.D
{\bf61}, 074018 (2000).\\
10. I.Balitsky,  Nucl.Phys.B {\bf463}, 99 (1996);  Phys.Lett.B
{\bf518}, 235 (2001).\\
11. D.A.Fagundes and M.V.T.Machado, Phys.Rev.D {\bf107}, 014004
(2023).\\
12. Qing-Dong Wu et al., Chin.Phys.Lett. {\bf33}, 012502 (2016).\\
13. M.Genovese, N.N.Nikolaev and B.G.Zakharov, J.Exp.Theor.Phys.
{\bf81}, 633 (1995).\\
14. I.P.Ivanov, N.N.Nikolaev, Phys.Rev.D {\bf65}, 054004 (2002).\\
15. N.N.Nikolaev and V.R.Zoller,  Phys.Lett.B {\bf509}, 283
(2001).\\
16. N.N.Nikolaev, W.Schafer, B.G.Zakharov and V.R. Zoller,
J.Exp.Theor.Phys. Letters {\bf84}, 537 (2007).\\
17. A.Accardi et al., Eur.Phys.J.A {\bf52}, 268 (2016).\\
18. D.P.Anderle et al., "Electron-Ion Collider in China," Frontiers of Physics {\bf16}, 64701  (2021).\\
19. J.Rausch, V.Guzey and M.Klasen, Phys.Rev.D {\bf107}, 054003 (2023).\\
20. H.Khanpour and S.Atashbar Tehrani, Phys.Rev.D {\bf93}, 014026
(2016).\\
21. H.Khanpour et al., Phys.Rev.D {\bf104}, 034010
(2021).\\
22. G.R.Boroun and B.Rezaei, arXiv:2303.07654; G.R.Boroun,
B.Rezaei and F.Abdi, arXiv:2305.01893.\\
23. F.Carvalho, F.O.Duraes, F.S.Navarra and S.Szpigel, Phys.Rev.C
{\bf79}, 035211
(2009).\\
24. J.Raufeisen, Acta Phys.Polon. B {\bf36}, 235 (2005).\\
25. Nestor Armesto, Carlos A. Salgado, Urs Achim Wiedemann,
Phys.Rev.Lett. {\bf94}, 022002 (2005).\\
26. N.Armesto, Eur.Phys.J.C 26, 35 (2002).\\
27. C.Marquet, Manoel R.Moldes and P.Zurita, Phys.Lett.B {\bf772},
607 (2017).\\
28. M.A.Betemps and M.V.T.Machado, Eur.Phys.J.C {\bf65}, 427
(2010).\\
29. E.R.Cazaroto, F.Carvalho, V.P.Goncalves and F.S.Navarra,
Phys.Lett.B {\bf671} , 233(2009).\\
30. F.Muhammadi and B.Rezaei, Phys.Rev.C {\bf106}, 025203
(2022).\\
31. N.Armesto, C.Merino, G.Parente and E.Zas,
Phys.Rev.D {\bf77}, 013001 (2008).\\
32. K.Golec-Biernat  and M.Wusthoff, Phys.Rev.D {\bf59},
014017 (1998); Phys.Rev.D {\bf60} 114023 (1999).\\
33. K. Golec-Biernat and S.Sapeta, J.High Energy Phys. {\bf03},
102 (2018).\\
34. J.Bartels, K.Golec-Biernat and H.Kowalski, Phys. Rev.
D{\bf66},
014001 (2002).\\
35. B.Blaettel, G.Baym, L.L.Frankfurt and M.Strikman,
Phys.Rev.Lett. {\bf70}, 896 (1993); L.Frankfurt, A.Radyushkin and
M.Strikman, Phys.Rev.D {\bf55}, 98 (1997).\\
36. G.R.Boroun and B.Rezaei, arXiv:2309.04832.\\
37. A.V.Kotikov, A.V.Lipatov and P.Zhang, Phys.Rev.D {\bf104}, 054042 (2021).\\
38. J.L.Albacete, N.Armesto, J.G.Milhano, C.A.Salgado, and
U.A.Wiedemann, Eur.Phys.J.C {\bf43}, 353 (2005).\\
39. Samuel S.M.Wong, Introductory Nuclear Physics, Prentice-Hall
of India Private Limited (2007).\\
40. D.P.Anderle et al., Front.Phys. {\bf16}, 64701 (2021).\\
\end{document}